\documentclass[%
reprint,
twocolumn,
showpacs,preprintnumbers,
amsmath,amssymb,
aip,
nofootinbib,
longbibliography, 
floatfix,
]{revtex4-2}

\usepackage{units}

\usepackage{upgreek}
\usepackage{textcomp}
\usepackage{enumitem}  
\usepackage{graphicx}
\usepackage{epstopdf}
\usepackage{dcolumn}
\usepackage{bm}
\usepackage{hyperref}

\usepackage{xcolor}
\definecolor{mylinkcolor}{rgb}{0.0,0.0,0.66}
\hypersetup{colorlinks=true,linkcolor=mylinkcolor,urlcolor=mylinkcolor,citecolor=mylinkcolor,filecolor=mylinkcolor,breaklinks=true,linktocpage=true,linktoc=all}

\usepackage[normalem]{ulem}
\usepackage{color,soul} 
\usepackage{lipsum}  
\usepackage[export]{adjustbox}

\newcommand{\ie}{\textit{i.e.,}\ }

\newcommand{\sdist}{\kern 0.20em}
\renewcommand{\eqref}[1]{Eq.\sdist(\ref{#1})}

\mathchardef\mhyphen="2D

\usepackage{color}

\emergencystretch=\maxdimen
\begin{document}


\title{Collective plasma effects of electron-positron pairs in beam-driven QED cascades}
\author{Kenan Qu}
\email[Corresponding author, ]{kq@princeton.edu}
\affiliation{Department of Astrophysical Sciences, Princeton University,  Princeton, New Jersey 08544, USA \looseness=-1 }  
\author{Sebastian Meuren}
\affiliation{Department of Astrophysical Sciences, Princeton University,  Princeton, New Jersey 08544, USA \looseness=-1 }  
\affiliation{Stanford PULSE Institute, SLAC National Accelerator Laboratory, Menlo Park, California 94025, USA \looseness=-1 } 
\author{Nathaniel J. Fisch}
\affiliation{Department of Astrophysical Sciences, Princeton University,  Princeton, New Jersey 08544, USA \looseness=-1 }

\date{\today}

\begin{abstract}
	Understanding the interplay of strong-field QED and collective plasma effects is important for explaining extreme astrophysical environments like magnetars. 
	It has been shown that QED pair plasma can be produced  and observed by passing a relativistic electron beam through an intense laser field. This paper presents in detail multiple sets of 3D QED-PIC simulations to show  the creation of pair plasma in the QED cascade. The beam driven method enables a high pair particle density and also a low particle Lorentz factor, which both play equal roles on exhibiting large collective plasma effects. 	Finite laser frequency upshift is observed with both ideal parameters ($\unit[24]{PW}$ laser laser colliding with $\unit[300]{GeV}$ electron beam) and with existing technologies ($\unit[3]{PW}$ laser laser colliding with $\unit[30]{GeV}$ electron beam). 
\end{abstract}


\maketitle


\section{Introduction}

When the Schwinger field~\cite{Schwinger_1951} is greatly exceeded, it is possible to reach the so-called {\it QED plasma regime} which exhibits both strong-field quantum effects as well as  collective plasma effects. 
The QED plasma regime is characterized by both high field strength to produce pairs and relatively high pair density to exhibit the collective effects.  Above the QED critical
limit $\unit[10^{18}]{Vm^{-1}}$, photons and electron-positron pairs are created in a cascaded manner~\cite{di_piazza_extremely_2012, sokolov_pair_2010, hu_complete_2010, thomas_strong_2012, neitz_stochasticity_2013, bulanov_electromagnetic_2013, blackburn_quantum_2014, green_transverse_2014, vranic_all-optical_2014, blackburn_scaling_2017, vranic_multi-gev_2018, magnusson_laser-particle_2019}. 
The resultant pairs travel at relativistic speeds, usually with  high Lorentz factors. 

This extreme regime is reached in intriguing astrophysical environments like magnetars~\cite{kaspi_magnetars_2017, cerutti_electrodynamics_2017}, binary neutron-star mergers~\cite{xue_magnetar-powered_2019, price_producing_2006}, and core-collapse supernovae explosions~\cite{mosta_large-scale_2015, akiyama_magnetorotational_2003}. 
For example, magnetars~\cite{lin_stringent_2020, ridnaia_peculiar_2020, li_identification_2020, bochenek_fast_2020, collaboration_bright_2020} are filled with strong-field QED cascades of relativistic electron-positron pair plasma~\cite{chen_filling_2020,timokhin_maximum_2019, gueroult_determining_2019, melrose_pulsar_2016} in their magnetospheres. 
The relativistic particle emission in the varying magnetic field of magnetars is very likely responsible for the Fast Radio Bursts~\cite{marcote_repeating_2020, amiri_second_2019,ravi_fast_2019,bannister_single_2019}.  It  often turns out to be of paramount importance~\cite{uzdensky_plasma_2014, uzdensky_extreme_2019, zhang_relativistic_2020} to appreciate the collective plasma effects in these extreme  
environments~\cite{wang_testing_2020,abbott_gw190425_2020, ligo_scientific_collaboration_and_virgo_collaboration_gw170817_2017, palenzuela_electromagnetic_2013,anderson_magnetized_2008}. 

To describe these collective effects, it might be surmised that the full machinery of collective plasma effects can simply be carried over to the QED pair-plasma regime.  It also might be surmised that, if the dynamics underlying these collective effects can be considered understood, then there would be few surprises in predicting collective phenomena.  
However, without probing this regime in laboratory experiments, there cannot really be sureity that there will be no such surprises.  
After all, a pair plasma, created out of extreme field energy, and constantly subject to both creation and recombination, is an exotic state of matter.  Can we be sure that the created particles obey the same collective effects found in less exotic plasmas?  
It behooves us to check experimentally the theoretical expectations and simplifications.
Hence, given the high interest in this regime, and given the high intensity lasers that are now available, it is now both critical and timely to perform the laboratory experiments that might exhibit the expected QED collective effects.

However, realizing the QED plasma regime -- and probing its collective effects -- while technically possible, is not so simple.  The QED plasma dynamics have been explored theoretically ~\cite{amaro2021optimal, Luo_scrp2018, Liu_PoP_2018, Guo_PPCF2019, Martinez_NJP2018, Luo_PPCF2018, Zhang_PoP2021, Capdessus_PoP2020, Yu_PPCF2018, Lecz_PPCF2019, Gu_MRE2019, Yu_PRL2019, Ong_PoP2019, Blackburn_2020, Zhang_PRA2019, Lobet_2017, Gong_PRAB2019, Li_PRL2020, Slade_NJP2019, Gong_PPCF2018, Shi_PRE2018, Samsonov_MRE2021}.  Possibilities for creating the QED regime in laboratory  experiments have also been explored~\cite{bell_possibility_2008,fedotov_limitations_2010, bulanov_schwinger_2010, nerush_laser_2011, elkina_qed_2011,jirka_electron_2016, grismayer_laser_2016, zhu_dense_2016,tamburini_laser-pulse-shape_2017,gonoskov_ultrabright_2017, grismayer_seeded_2017, savin_energy_2019, Qu_QED2021}.
Note that the particle density necessary to manifest collective plasma effect depends on the observation method. 
One way is to measure the wavelength of the plasma emission driven by a laser field. 
To emit at an infrared wavelength (to which the detectors are most sensitive), the particle density needs to reach near $\unit[10^{25} \sim10^{27}]{m^{-3}}$. This corresponds to a $\unit[0.001]{nC} \sim \unit[0.1]{nC}$ pair plasma in a $\unit[1]{\mu m^{-3}}$ volume sphere.

The all-optical approach to reaching and observing the QED regime employs two colliding ultra-strong lasers. 
The strong beat wave accelerates the seed electrons to  relativistic speeds, which in turn boosts the laser field amplitude. 
As soon as the QED critical field is reached in the electron rest frame, high energy photons are emitted and pairs are created. 
The strong laser continues to accelerate the particles to induce a QED cascade. 
With a sufficiently strong laser field, a QED cascade is produced in the rest frame of the pair particles leading to a pair plasma. 
It was proposed~\cite{bell_possibility_2008} that laser  intensities above $\unit[10^{24}]{Wcm^{-2}}$ are sufficient to probe the QED critical field in the pair particle rest frame.  
This all-optical method has prompted investigation both analytically and numerically~\cite{bell_possibility_2008, fedotov_limitations_2010, bulanov_schwinger_2010, nerush_laser_2011,elkina_qed_2011, jirka_electron_2016, grismayer_laser_2016,zhu_dense_2016,tamburini_laser-pulse-shape_2017, gonoskov_ultrabright_2017, grismayer_seeded_2017,savin_energy_2019}. 
However, the $\unit[10^{24}]{Wcm^{-2}}$ laser intensity required by the all-optical method needs a technologically challenging tight focus of a $\unit[100]{PW}$ laser~\cite{cartlidge_light_2018, OPCPA_2019, danson_2019}. 
Solving this challenge will depend on substantial development beyond current state-of-the-art laser technology. 
Even if the pair plasma is created, the pair particles have a large Lorentz factor of $\sim10^3$.  The large Lorentz factor is important, since the  pairs  contribute to the plasma frequency (observed in the laboratory frame) inversely to their relativistic mass. The smallness of this contribution means that even higher densities of pairs are necessary to make substantive contributions to collective plasma effects.

Here we describe an alternative, less technologically challenging than the all-optical approach to reaching the QED plasma regime by colliding a less intense laser pulse with an electron beam. 
The idea, introduced recently~\cite{Qu_QED2021} and expanded upon here, is to generate a quasi-neutral pair-plasma with a density that is comparable to the critical one by using the combination of a $\unit[30]{PW}$ laser and a {dense} $\unit[300]{GeV}$ electron beam or by using less stringent parameters. This method circumvents the technological limitations by taking advantage of the high quality energetic electron beam facilities to boost the laser intensity in the particle rest frame. 
It also facilitates observation of the QED collective effects, since in the end, what must be solved is not merely reaching the QED plasma regime, but reaching the QED plasma regime {\it and} observing its collective effects.  
Hence, what must be optimised is the {\it joint production-observation problem}.
If the pairs produced are less energetic, then, at much lower pair densities, there can be larger, more easily observed, contributions to collective effects.

This paper is organized as follows: In Sec.~II, we briefly outline why the beam-laser approach directly engages the joint production-observation problem. 
In Sec.~III, in order to provide context for our numerical simulations, we review the physics of the QED cascade in an electron-beam-laser collision.  In Sec.~IV, we introduce the setup of our 3D PIC QED simulations and show the production of electron-positron pairs. In Sec.~V, we focus on the pair deceleration through both synchrotron emission and  laser radiation pressure. The important pair reflection condition is introduced here. In Sec.~VI, we analyze the laser dynamics and show how its spectrum changes with the increasing plasma frequency. Multiple optical detection methods are explained including laser central frequency shift, chirping, and homodyne detection of the laser phases. In Sec.~VII, we verify the scaling of the laser frequency shift with different electron beam and laser parameters. In Sec.~VIII, we demonstrate the possibility of creating and observing a QED plasma using state-of-the-art parameters, \ie a $\unit[3]{PW}$ laser and a dense $\unit[30]{GeV}$ electron beam (through, for example, a combination of FACET-II with LCLS-Cu RF LINAC~\cite{FACET-II}). In Sec.~IX, we present our conclusions.

\section{Beam-laser approach to the production-observation problem}



Consider the rest frame of high energy particle beams, because reaching the Schwinger field limit directly in laboratories is still beyond the capability of current technology. 
A multi-GeV electron beam from a particle accelerator can have a Lorentz factor of over $10^4$, which can boost the laser field by the same number. The seminal SLAC E-144 experiment~\cite{bula_observation_1996,burke_positron_1997} has already used this method to observe evidence of nonlinear Compton scattering and Breit-Wheeler pair production by colliding a ${\sim\unit[10^{18}]{Wcm^{-2}}}$  laser and a $\sim\unit[50]{GeV}$ electron beam. Due to the relatively low laser intensity, only a limited number of positrons were produced, so that collective plasma effects could not be observed. The upcoming experiment SLAC FACET-II will deploy a new laser with over $\unit[10^{20}]{Wcm^{-2}}$ peak intensity. Combined with the LCLS-Cu LINAC~\cite{white_2020, white_ultra_2018, yakimenkoPRL19}, a pair multiplication factor over unity can be achieved providing a unique opportunity to explore the QED pair plasma.

Compared with the all-optical method, the beam-driven approach lowers the laser intensity requirement by two orders of magnitude. This is because  particle accelerators can produce multi-GeV or even tens of GeV electron beam energy, corresponding to Lorentz factors of $10^4\sim10^5$. A PW-level laser can already induce QED pair multiplication. Such laser systems are routinely operated in several laboratories~\cite{danson_2019}. 
The all-optical method only accelerates the electrons to a Lorentz factor similar to the laser dimensionless amplitude $a$; Even reaching $\gamma\sim 10^3$ needs laser intensity of over $\unit[10^{24}]{Wcm^{-2}}$ ~\cite{bell_possibility_2008,fedotov_limitations_2010, bulanov_schwinger_2010, nerush_laser_2011,elkina_qed_2011,jirka_electron_2016, grismayer_laser_2016,zhu_dense_2016,tamburini_laser-pulse-shape_2017,gonoskov_ultrabright_2017, grismayer_seeded_2017,savin_energy_2019}  and thus requires large $\unit[100]{PW}$-scale laser facilities~\cite{cartlidge_light_2018, OPCPA_2019, danson_2019}.  


As the pair density grows in a QED cascade, collective plasma effects emerge. 
Existing experimental detectors like magnetic spectrometers can distinguish electrons and positrons, 
but cannot measure the pair density or observe a collective plasma effect. 
The figure of merit for collective plasma dynamics is determined by the plasma frequency ($\omega_p$) which is proportional to the ratio of pair particle density ($n_p$) and pair Lorentz factor ($\gamma$), \ie $\omega_p \propto \sqrt{n_p/\gamma}$. 
Since the average pair Lorentz factor is similar to the laser dimensionless amplitude, the lower laser intensity needed for the beam-driven approach  greatly reduces the average Lorentz factor of the produced pair plasma. 
It thus exhibits higher plasma frequency even if the colliding lasers can produce the same plasma density. 
The counterpropagating configuration of the laser pulse and pairs also further slows down the pairs through laser radiation pressure. Thus, the pair plasma created in a beam-driven QED cascade is easier to detect than one created with the all-optical approach. 
All of these advantages, favoring the beam-driven approach over the all-optical approach,  accrue from the simple fact that the pair masses, and hence their contributions to the pair plasma frequency, are inversely proportional to the pair Lorentz factor. 

There are further advantages to the beam-driven approach.
The laser which is used to create the pair plasma also informs the pair plasma property through its change of spectrum. Both as the pair plasma forms and as it slows down, the plasma frequency increases inside the laser field. The increase of plasma frequency abruptly reduces the vacuum refractive index which mediates the laser. The consequence is that the laser frequency is upshifted and the laser wavelength is blue shifted, according to the theory of temporal change of optical refractive index~\cite{wilks1988frequency, esarey_frequency_1990, PRL_Wood_1991, Kenan_slow_ionization, Shvets2017,Nishida2012, Kenan_2018_upshift,Edwards_Chirped,Bulanov2005, Peng_PRApp2021}. 
The optical emission thus serves as a robust signature of the creation of collective plasma effects in the QED plasma despite the small plasma volume and relativistic plasma motion. 
In fact, the small plasma volume ($\mu$m-scale)  eliminates the possibility of using conventional detection methods, e.g., by observing plasma instabilities like the two-stream instability~\cite{Greaves_prl_1995}, the Weibel instability~\cite{Fried_1959}, or stimulated Brillouin scattering~(SBS)~\cite{Edwards_prl_2016}.


\section{Pair generation through beam-driven QED cascade}

To give context to numerical simulations that support the beam-driven approach, we provide first a brief overview of the QED cascade process. 
In an electron-beam-driven QED cascade, an energetic electron beam collides with a counter-propagating strong laser field. The laser amplitude is greatly boosted to exceed the critical field $E_s$ in the electron rest frame. Thus, the electrons emit photons which further split into electron-positron pairs. Each of the pair particles, given sufficient energy, continues the photon emission and pair generation process to induce a cascaded generation of pairs. The process converts high electron beam energy into large pair numbers. In this section, we lay out the properties of beam-driven QED cascade while we briefly explain how the electron-positron pairs are generated.

In the initial stage of the collision, the electron beam has the maximum energy with a Lorentz factor $\gamma$. It boosts the counter-propagating laser field by the same factor $\gamma$ in the rest frame of the electrons. Such a boost aims to produce a large quantum parameter for electrons $\chi_e \equiv E^*/E_s = \gamma \big|\bm{E}_\perp + \bm{\beta}\times c\bm{B} \big| / E_s$, where $\bm{E}$ and $\bm{B}$ are the electric and magnetic field of the laser in the laboratory frame, and $\beta$ is the electron beam speed normalized to the speed of light $c$. The strong laser field drives the electron motion, causing emission of photons~\cite{di_piazza_extremely_2012, landau_quantum_1981}.  

The photon emission spectra differ depending on the quantum parameter $\chi_e$. The regime $\chi_e \gtrsim 1$ is reached near the laser intensity peaks where quantum synchrotron emission causes the electrons to emit almost all of its energy into a single gamma ray. In the regions of low laser intensity $\chi_e \ll 1$, the emission is classical, and the emitted photon $\hbar\omega $ only takes a small portion of the electron energy $\mathcal{E}$, \ie $\hbar\omega \sim \chi_e \mathcal{E}$. 

The low energy photons would escape the laser focus spot without decaying into pairs. 
But the high-energy gamma ray photons are highly likely to decay into an electron-positron pair in the same strong laser field. For the interest of this paper, we focus on the Breit-Wheeler process that a photon decays into one pair of electron and positron. The decaying process  depends on the quantum parameter of the emitted photon $\chi_\gamma \equiv [\hbar\omega/(2m_ec^2)] \big|\bm{E}_\perp + \hat{\bm{k}}\times c\bm{B}\big| / E_s$ with $\hbar\omega$($\hbar\bm{k}$) being the photon energy (momentum), and $m_e$ being the electron rest mass. The pair generation happens only when $\chi_\gamma$ is above the unit threshold value. For $\chi_\gamma \gg 1$, the photon transfers almost all of its energy to either the electron or positron, whiles for smaller $\chi_\gamma$ values the photon energy is more symmetrically partitioned.

Therefore, a very intense laser can cause a cascade of gamma rays and pairs from a single energetic electron. Each subsequent emission and decay process transfers the energy  predominantly into one new particle and creates many other particles with lower energies. The total energy of the pairs and photons are conserved, which allows us to find the pair number multiplication factor. Combined with the fact that the gamma photon decay process terminates when $\chi_\gamma <1$, the pair number grows by a factor that is approximately equal to the maximum quantum factor $\chi_e$ at the laser peak amplitude. Thus, for an electron beam with original Lorentz factor $\gamma_0$ and density $n_e$ and a laser with dimensionless amplitude $a_0 \equiv eE/(m_ec^2\omega_0)$ and frequency $\omega_0$, the pair number multiplies as 
\begin{equation} \label{multip}
	n_p \sim \widetilde{\chi}_e n_e, \qquad  \widetilde{\chi}_e \approx 2a_0\gamma_0 (\hbar\omega_0)/(m_ec^2).
\end{equation}
This relation holds only if the laser pulse is sufficiently wide and long. Finite laser pulse waist and duration could cause deviation of the pair multiplication factor from the estimation, but the linear scaling should nevertheless hold in general.  It should be born in mind that the quantum parameter for the pairs $\chi_e$ and for the photons $\chi_\gamma$ depend on the local laser field strength, and hence they vary at a different location and time. The pair number multiplication factor $\widetilde{\chi}_e$ is approximately equal to the maximum  quantum parameter for the pairs, and hence is a fixed value for certain laser pulse and electron beam energy. 

The quantum photon emission process has a lower requirement for the field and it terminates when $\chi_e \ll1$. Though the relatively low energy photons do not decay into pairs, they play an important role in decelerating the pair particles, which contributes to higher collective plasma effects. 

Since the pair formation rate $t_f^{-1} \sim 2a_0 \omega_0$ is proportional to the laser amplitude $a_0(\gg1)$, the pairs are more likely to be created in the region where the laser field is strong. The strong laser field drives pair oscillation immediately when they are generated. The collective effects, if they were to be probed, are manifested through the oscillation of the pair particles in the strong laser field.


\section{3D PIC QED simulations of pair creation}

\begin{figure}[ht]
	\centering
	\includegraphics[width=\linewidth,valign=t]{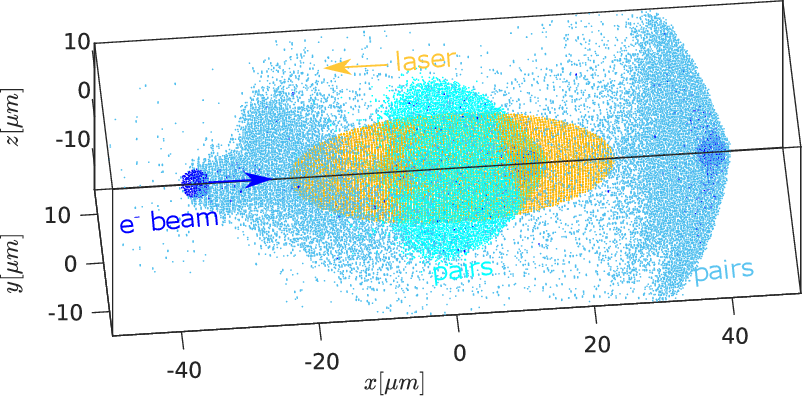}
	\caption{Schematics of the 3D QED-PIC simulation of an energetic electron beam (deep blue) colliding with a multi-PW laser pulse (yellow) to create an electron-positron pair plasma. The volume of the pair plasma at different times is denoted as green (${t=0.21}$ps) and light blue (${t=0.3}$ps) dots. } 
	\label{sch}
\end{figure}

The above analysis shows that an electron-beam-driven QED cascade requires including an intense laser field with $a_0\gg1$ and an electron beam with high $\gamma$ factor such that $\chi_e\gtrsim1$. Towards this limit, our simulations consider head-on collision of a $1$~nC electron beam of $300$~GeV~\cite{CLIC,AWAKE}, shown as a blue sphere in Fig.~\ref{sch}, and a  $24$~PW laser pulse~\cite{OPCPA_2019} with wavelength $\lambda=\unit[0.8]{\mu m}$, shown as an yellow spheroid. The corresponding dimensionless laser amplitude is $a_0 \approx 170$ and the maximum quantum parameter is $\widetilde{\chi}_e \approx 220$ at the Gaussian waist in the focal plane, and $\widetilde{\chi}_e \approx 600$ at the laser focus.

\begin{figure}[h]
	\centering
	\includegraphics[width=\linewidth,valign=t]{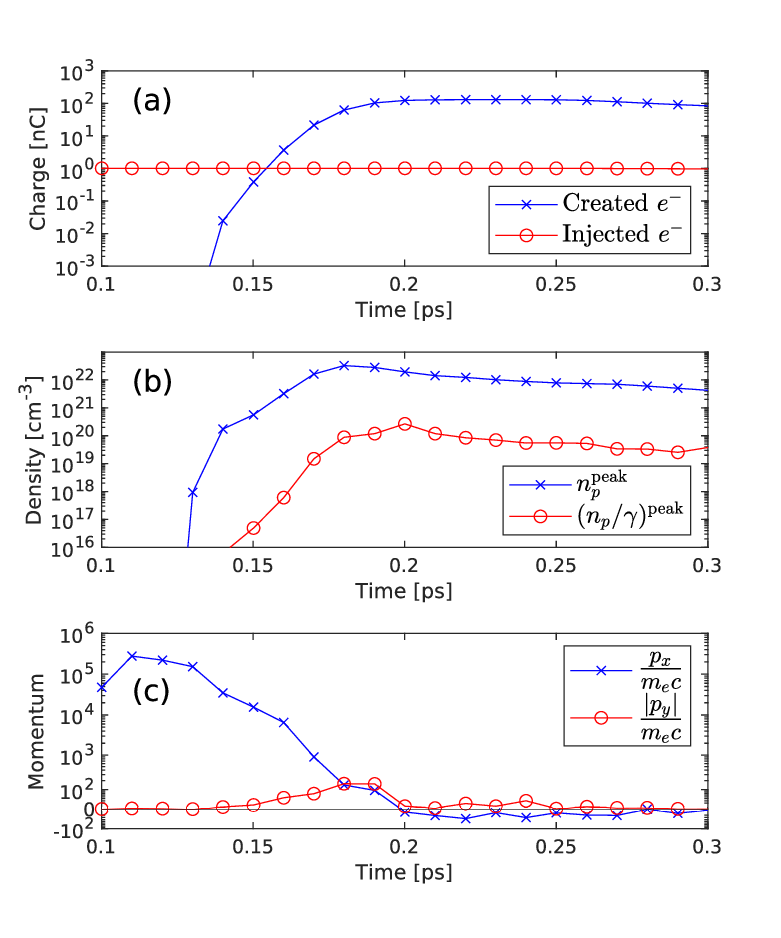}
	\caption{(a) Evolution of total charges of the injected electrons (red) and created electrons (blue). (b) Evolution of the peak pair plasma density $n_p$ (blue), and the parameter $n_p/\gamma$ (red) which determines the laser frequency upshift. (c) Evolution of the pair particle momenta in the longitudinal (blue) and transverse (red) directions, normalized to $m_ec$. The local particle density $n_p$, Lorentz factor $\gamma$, and momentum $p_{x,y}$ are denoted by their respective averaged values in a single simulation cell. } 
	\label{evo}
\end{figure}

The electron beam has a spherical Gaussian number density $n_e(\bm{r}) = n_{e0} \exp[-r^2/(2r_0^2)]$, where $n_{e0} = 4\times10^{20} \,\mathrm{cm}^{-3}$ is the peak density and $r_0 = 1\mu$m is the rms radius of the sphere. The counter-propagating laser pulse is linearly polarized in the $y$ direction and propagates in the $-x$ direction. It has a Gaussian distribution in both transverse and longitudinal directions with $I = I_0 \cdot[w_0/w(x)]^2 \exp[-2\rho^2/w^2(x)] \exp[-2t^2/\tau^2]$ where $I_0 = 6\times10^{22}\, \mathrm{Wcm}^{-2}$ is the peak intensity, $w_0 = 5\mu$m is the waist at $x=0$, $w(x)=w_0\sqrt{1+(x/x_R)^2}$, $x_R=\pi w_0^2/\lambda \approx \unit[98]{\mu m}$ is the Rayleigh length, and $\tau=\unit[50]{fs}$ is the pulse duration (intensity FWHM: $\sqrt{2\log(2)}\tau\approx\unit[59]{fs}$, electrical field FWHM: $2\sqrt{\log(2)}\tau\approx\unit[83]{fs}$). Each dot in Fig.~\ref{sch} represents a region with pair density above $\unit[2\times10^{20}]{cm^{-3}}$ or laser intensity above $\unit[5\times 10^{20}] {Wcm^{-2}}$. The simulation starts at $t=\unit[-0.205]{ps}$ and the center of laser pulse and electron beam arrive at their corresponding boundaries at $t\simeq 0$. The simulation terminates at $t=\unit[0.32]{ps}$ when the laser pulse exits the simulation box.

We choose a linearly polarized laser for the simulation because it can achieve a larger pair multiplication factor compared to using a circularly polarized laser at the same energy. The higher pair multiplication origins from the exponential dependence of the pair growth rate on the field amplitude. A linear polarized laser has $\sqrt{2}$-fold higher peak field amplitude than a circularly polarized laser at the same laser energy. When averaging over the laser period, the higher peak amplitude leads to a larger number of created pairs. 

\begin{figure}[th]
	\centering
	\includegraphics[width=\linewidth,valign=t]{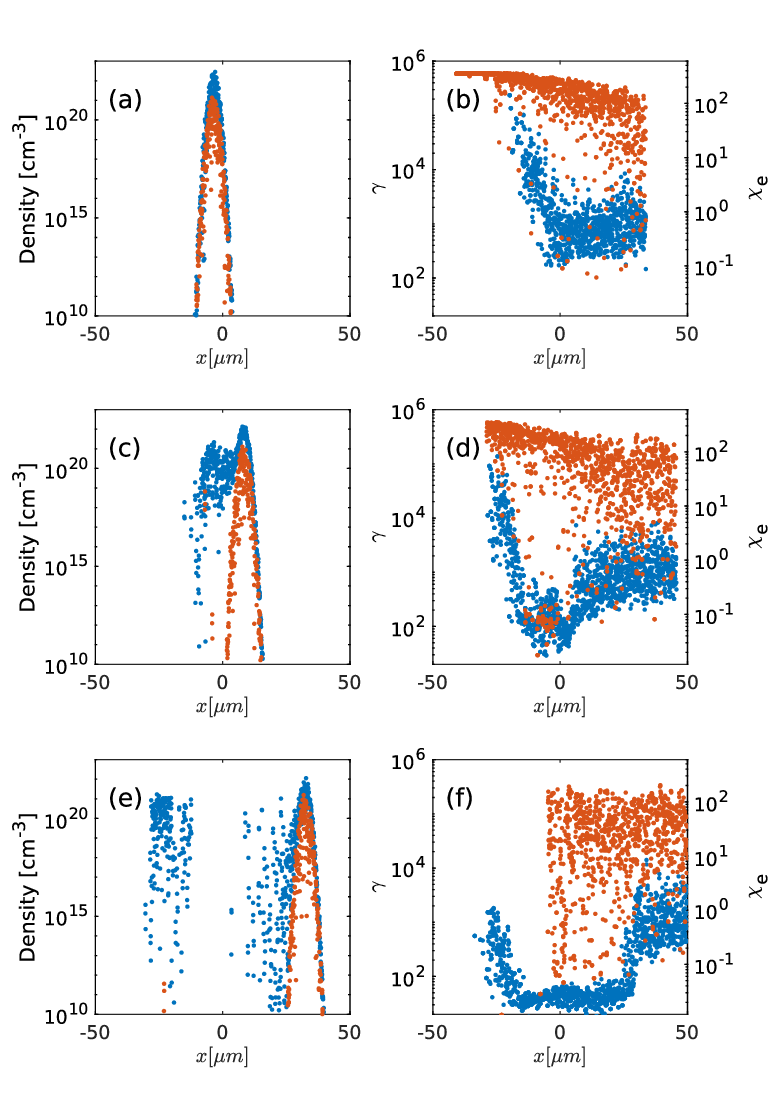}
	\caption{ The density (left column) and Lorentz factor $\gamma$ (right column) of the injected electrons (red) and created electrons (blue) at line $y=z=0$. The electron quantum factor $\chi_e$ is calculated at the peak laser amplitude $a_0=170$. The three rows show the snapshots at $\unit[0.16]{ps}$ (a-b), $\unit[0.2]{ps}$ (c-d), and $\unit[0.28]{ps}$ (e-f), respectively.  } 
	\label{dens}
\end{figure}

\begin{figure}[th]
	\centering
	\includegraphics[width=\linewidth,valign=t]{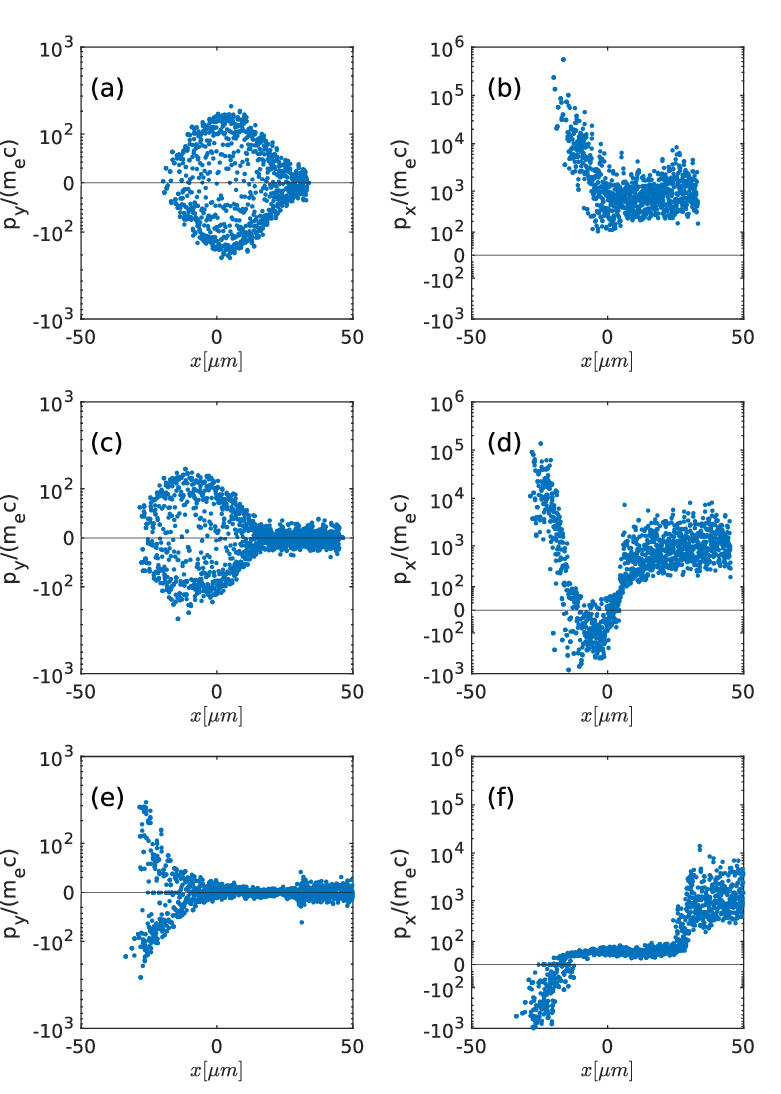}
	\caption{ The transverse $p_y$ (right column) and longitudinal $p_x$ (left column)  momenta of the created electrons  at line $y=z=0$, normalized $m_e c$. The three rows show the snapshots at $\unit[0.16]{ps}$ (a-b), $\unit[0.2]{ps}$ (c-d), and $\unit[0.28]{ps}$ (e-f), respectively.  } 
	\label{momen}
\end{figure}

The simulations were performed using the PIC code EPOCH~\cite{EPOCH2015,RIDGERS2014273}  with the QED module~\cite{savin_energy_2019,blackburn_scaling_2017,jirka_electron_2016}. The simulation box measured $100\mu\mathrm{m} \times 30\mu\mathrm{m} \times 30\mu\mathrm{m} $ is discretized into $4000 \times 300 \times 300$ cells. The charged particles are represented by near $6\times10^8$ computing particles. The time step is determined by both the Courant–Friedrichs–Lewy condition and the inverse plasma frequency, \ie it is chosen as the smaller value of the minimum plasma oscillation period in any cells and the smallest cell dimension multiplied by $0.95/c$. The actual time step in our simulations is $\unit[0.083]{fs}$, which is well below the maximum possible photon emission time~\cite{RIDGERS2014273} $t_f \gtrsim0.36$ fs in all our simulations. 

With the large quantum parameter, the collision quickly creates a pair plasma with an increasing charge number. Figure~\ref{evo}(a) shows the evolution of total charge of the injected electrons (red circles) and created electrons (blue crosses). The  injected electron beam remains $\unit[1]{nC}$ throughout the interaction. The pair electron charge grows exponentially until reaching $\unit[139]{nC}$ total charge at $\sim\unit[0.2]{ps}$ and then remains unchanged afterwards. 

The peak density of the created pairs, shown as the blue curve with cross markers in Fig.~\ref{evo}(b), quickly grows to a peak value of $n_p= 82n_{e0} = 3.28\times 10^{22} \,\mathrm{cm}^{-3}$, but begins to slowly decrease at $\unit[0.17]{ps}$. The decrease of peak density is caused by plasma volume expansion, illustrated at three different stages in Fig.~\ref{sch}. The red curve in Fig.~\ref{evo}(b) shows the parameter $n_p/\gamma$ which determines the plasma frequency, which we will explain in detail in the next section. 
Since the pair particles are mostly created in the region of strong laser field, the pair particles immediate accelerate transversely causing volume expansion. The transverse motion, shown as the growing transverse momentum in  Fig.~\ref{evo}(c), also allows the particle to escape the high intensity laser focus resulting in a lower total charge than predicted by Eq.~(\ref{multip}). The blue curve with cross markers in Fig.~\ref{evo}(c)  shows the change of pair moving direction in the later stage of cascade, and we will explain it in detail in the next section.

To get more insight into the dynamics of pair generation, we analyze the pair density and momentum on the center line $y=z=0$ with peak laser intensity. Each data point on the line is plotted as a dot in Fig.~\ref{dens} and Fig.~\ref{momen}. 
The variances of the pair density and energy in the plots are caused by the stochastic property of the QED process. 
The top row of these two figures show the snapshot at $\unit[0.16]{ps}$. At this time, the center of the electron sphere has not reached the laser focal plane though a significant amount of pairs have already been generated. The density plot in Fig.~\ref{dens}(a) shows that the generated pairs are limited to the region near the electron beam. The Lorentz factor plot in Fig.~\ref{dens}(b) shows that the electron beam energy is decreased by $2\sim3$ orders of magnitude from its initial value $\gamma_0=6\times10^5$ after passing through the laser peak. The generated pairs have an energy of near $10^3m_e c^2$, corresponding a quantum parameter of $\chi_e\lesssim 1$, after passing through the laser peak. This is the lowest particle energy that can be efficiently generated via the photon decay process. Beyond this point, the gamma photons can no longer decay into pairs and the QED cascade terminates, which can be seen from the saturation of charge growth in Fig.~\ref{evo}(a). However, the pairs continue to lose energy as shown in the last two rows of Figs.~\ref{dens} and \ref{momen}. We will discuss them in detail in the following section.

\section{Pair deceleration and pair reflection}

We observed and explained in the previous section that the injected electron beam would emit high energy photons which decay into pair particles in a strong laser field until the photon energy (and hence the generated pair energy) decreases to $\chi_\gamma\lesssim 1$ when the QED cascade terminates. The pairs, however, continue to lose energy via synchrotron radiation, as shown in Fig.~\ref{evo}(c).

The photon emission is dominated by quantum synchrotron radiation when $\chi_e\gtrsim 0.1$. The decrease of pair energy can be seen in Fig.~\ref{dens}(d) and (f), which shows the snapshot immediately after the center of the electron sphere passes through the laser peak. The blue curve in Fig.~\ref{dens}(d) shows that the generated pair energy further decreases to $10^2m_ec^2$, corresponding to $\chi_e\sim0.1$, at the tail of the electron beam. Depending on the laser amplitude $a_0$, quantum synchrotron emission can reduce the pair Lorentz factor to 
\begin{gather}
	\label{eqn:gamma}
	\gamma \lesssim 0.1 \frac{\gamma_0}{\widetilde{\chi}_e} \approx \frac{0.05}{a_0} \frac{m_ec^2}{\hbar\omega_0}.
\end{gather}
Figure~\ref{dens}(c) shows that these low-energy pairs are created through the secondary generation from the daughter pairs, and thereby they tailgate the injected electron beam. 

Quantum synchrotron emission stops when the particles lose sufficient energy or the laser amplitude becomes low, \ie ${\chi_e \lesssim0.1}$. Then classical radiation emission begins to dominate~\cite{di_piazza_extremely_2012}. The pair particles wiggle in the laser field to radiate electromagnetically with negligible quantum contributions like recoil or spin. The strong laser field drives transverse motion of the pairs, evidenced in Fig~\ref{momen}(a,c) for that the transverse pair momentum $p_y$ is enveloped by the laser profile and that $|p_y|/(m_ec)=a_0$ locally. Due to the conservation of canonical momentum, each charged particle transfer the amount of $a_0^2m_ec/(4\gamma)$ longitudinal momentum to a counter-propagating laser field upon entering it. It means that particles can be stopped or even reflected by the strong laser field if they have sufficiently low longitudinal momentum, \ie $p_x \lesssim a_0^2m_ec/(4\gamma)$, or equivalently, $\gamma \sim a_0$~\cite{li_attosecond_2015}. By comparing this condition with \eqref{eqn:gamma}, we find the threshold laser intensity for particle reflection 
\begin{gather} \label{eqn:reflection}
	a_{0,\mathrm{th}} \gtrsim \sqrt{0.05 m_ec^2/(\hbar\omega_0)}.
\end{gather} 
For optical lasers with $\hbar\omega_0 \sim \unit[1]{eV}$, the threshold is approximately $a_{0,\mathrm{th}} \gtrsim 100$, corresponding to intensity $I_{\mathrm{th}} \gtrsim 10^{22} \mhyphen 10^{23}\, \mathrm{Wcm}^{-2}$. Particle reflection is shown in Fig.~\ref{evo}(c) and  Fig.~\ref{momen}(d,f) as the pair longitudinal momentum $p_x$ becomes negative near the laser peak. The reflected pair can also be observed in Fig.~\ref{sch}(c) in which the pairs (light blue dots) spread throughout the simulation box at $t=0.2$ ps.

The particle reflection threshold is very important for probing the collective pair effects because the pair particle mass reaches their minimum value as they stop longitudinally. Since plasma dynamics is manifested through plasma frequency $\omega_p$ which is proportional to $\sqrt{n_p/\gamma}$, achieving lower particle energy is equally important with achieving higher particle density. Thus, we plot the parameter $n_p/\gamma$ in Fig.~\ref{evo}(b). The red curve with circle markers shows that the parameter $n_p/\gamma$ continues to increase even after the pair density reaches its peak value $n_p= 3.28\times 10^{22} \,\mathrm{cm}^{-3}$ at $t=\unit[0.18]{ps}$. The pair momentum decreases to its minimum value at $t=\unit[0.2]{ps}$ when $n_p/\gamma$ reaches its peak value of $2.67\times10^{20}\,\mathrm{cm}^{-3}$ at $t=\unit[0.2]{ps}$. Thus, the beneficial synchrotron radiation, which keeps reducing the pair energy, outweighs the density decrease between $t=0.18\, \mathrm{ps}$ and $t=0.2\, \mathrm{ps}$ until finally the latter process dominates. 

While the strong laser field causes the pairs to lose longitudinal momentum $p_x$, it at the same time increases the transverse momentum $p_y$. The maximum value of $p_y$ is identical to local laser amplitude $a_0$ due to conservation of canonical momentum. Therefore, the minimum pair Lorentz factor is equal to the laser amplitude $\gamma=a_0$ provided that the particle reflection threshold condition Eq.~(\ref{eqn:reflection}) is satisfied. This is evident in Fig.~\ref{dens}(d) and (f) which shows a minimum Lorentz factor of $\gamma\lesssim100$. Note that the particle reflection may happen behind the laser peak if we consider the finite time of particle deceleration by synchrotron radiation. 

Thus, the final pair density is approximately the multiplication of the quantum nonlinear parameter and the initial electron beam density; the final pair Lorentz factor is approximately equal to the laser $a_0$ factor, \ie
\begin{gather}
	\label{eqn:ruleofthumb}
	n_p \sim \widetilde{\chi}_e n_e, \quad \gamma_f \sim a_0.
\end{gather}
These relations are valid if the laser is above the threshold intensity $I_\mathrm{th}$ for particle reflection [\eqref{eqn:reflection}], and if the interaction time is long enough such that the cascade reaches its asymptotic state. 


\begin{figure}[th]
	\centering
	\includegraphics[width=\linewidth,valign=t]{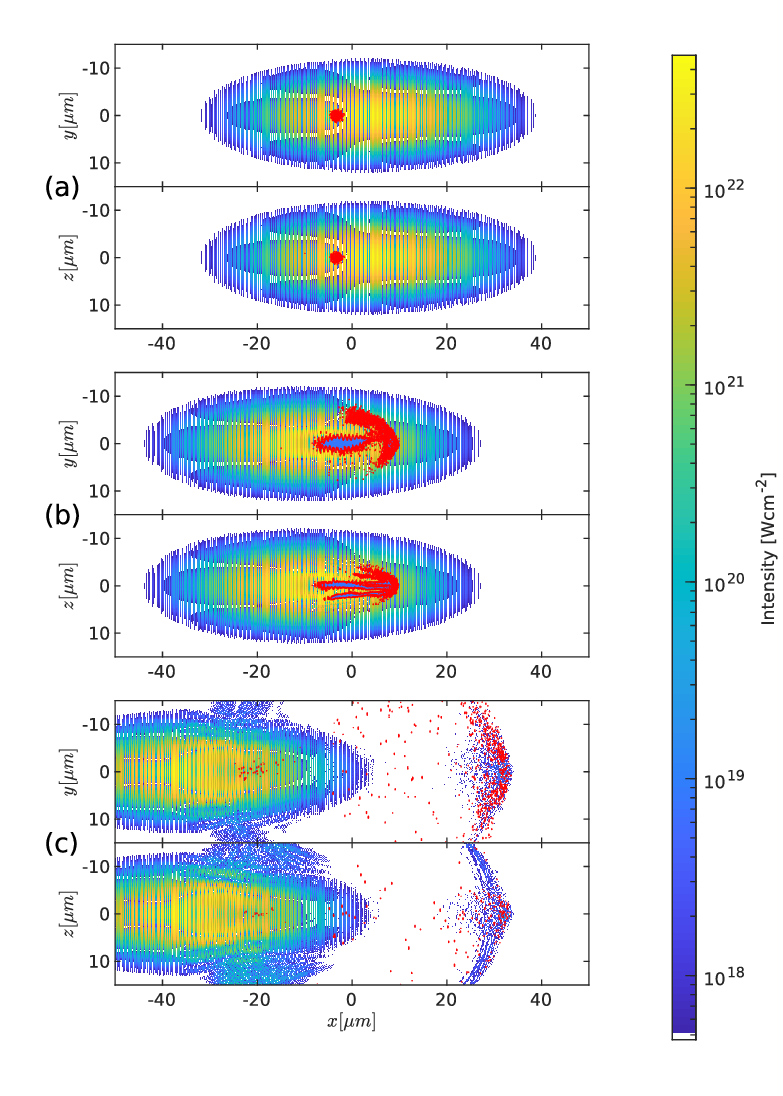}
	\caption{The two panels of each subplot show the laser profiles at the $z=0$ cross section and $y=0$ cross section, respectively. The red dots show the regions of $n_p/\gamma>\unit[1\times{}10^{19}]{{cm}^{-3}}$ at the corresponding planes. The snapshots are taken at $\unit[0.16]{ps}$ (a), $\unit[0.2]{ps}$ (b), and $\unit[0.28]{ps}$ (c), respectively} 
	\label{planes}
\end{figure}

To illustrate the creation of pairs more clearly, we plot the profiles of pair density and laser intensity in the $y=0$ plane and $z=0$ plane as the top and bottom panels of Fig.~\ref{planes}, respectively. Each red dot denotes a region with $n_p/\gamma>\unit[1\times{}10^{19}]{{cm}^{-3}}$ at the corresponding plane. It is seen in Fig.~\ref{planes}(a) that the pairs are initially created nearby the injected electron beam. They then expand to mainly the transverse directions shortly after the creation. Linear laser polarization breaks the cylindrical symmetry of pair motion. The $y$-polarized laser naturally accelerates the pairs more strongly in the $y$ direction than in the $z$ direction, which is seen in Fig.~\ref{planes}(b). The asymmetry of pair expansion increases as shown in Fig.~\ref{planes}(c). Figure~\ref{planes}(c) also reveals rich dynamics of the laser profile, which will be discussed in detail in the following section.

\section{Laser beam diffraction and  frequency upconversion}

Since the pair generation rate is proportional to the laser amplitude, the pairs are dominantly created near the peak laser field. The strong laser field thus drives the pairs into oscillation immediately after they are generated. The induced transverse current  radiates  electromagnetic fields. With non-negligible pair density, the radiation could reach a detectable level to reveal the pair dynamics. When the pair density reaches near the critical density, the radiation becomes strongly coupled to the input laser, causing a quantitative upshift of the laser frequency. Measuring the change of laser frequency allows us to unambiguously probe the collective pair plasma effects. 

Macroscopically, laser frequency upshift arises from non-adiabatic change of index of refraction, which determines the phase velocity of light. Suddenly created pairs reduces the index of refraction thereby increasing the laser phase velocity. It corresponds to an increased rate of local laser phase oscillation  and thus an upshift of laser frequency. 

Microscopically, the laser frequency upshift can be analyzed through finding the transverse current $\bm{J}_\perp$ of the pair particles. As the pairs are almost always generated when strong laser field is present, they are immediately driven into an oscillatory motion. Assuming that the pair particles have no transverse momentum at the time of generation. The laser field with vector potential $\bm{A}$ can transfer transverse momentum of $\bm{p}_\perp = e\bm{A}_\perp$ to the pairs. Thus, the pair transverse current is $\bm{J}_\perp = 2en_p \bm{p}_\perp/(\gamma m_e) = (2e^2 n_p/m_e) \bm{A}_\perp/\gamma = \varepsilon_0 \omega_p^2 \bm{A}_\perp$ with $\varepsilon_0$ being the permittivity of vacuum. Here, we identified the plasma frequency $\omega_p \equiv [2n_p e^2/(\gamma m_e \varepsilon_0) ]^{1/2}$. The factor of two accounts for the equal contribution of positrons and electrons to the laser dispersion relation. The transverse current couples to the laser field through the wave equation 
\begin{equation} \label{eqA}
	\nabla^2 \bm{A}_\perp -\frac{1}{c^2} \partial_t^2 \bm{A}_\perp = \frac{e^2}{m_ec^2\varepsilon_0} \frac{2n_p}{\gamma} \bm{A}_\perp \equiv \frac{\omega_p^2}{c^2} \bm{A}_\perp, 
\end{equation}
from which we see that a non-adiabatic change of plasma frequency  $\omega_p^2 \propto n_p/\gamma$ induces a change of laser frequency $\omega$. If the plasma frequency is small compared with the input laser frequency $\omega_p \ll \omega$, the laser frequency is approximated as 
\begin{equation}\label{frequp0}
	\omega \cong \omega_0 + \omega_p^2/(2\omega_0).
\end{equation}
If the plasma is created non-instantly, the change of laser frequency could be expressed in an integral form of the plasma frequency change at the retarded position $X=x+c(t-t')$:
\begin{equation}\label{frequp1}
	\omega(x,t) = \omega_0(x) + \frac{1}{2\omega_0} \int_0^t dt' \big[\partial_T \omega_p^2(X,T) \big] _{X = x+c(t-t')}^{T = t'}.
\end{equation}
The laser wave vector changes correspondingly obeying the dispersion relation 
\begin{multline}\label{kup1}
	k(x,t) \cong \omega(x,t)/c - \omega_p^2(x,t)/(2\omega_0c) \\
	= k_0(x) + \frac{1}{2\omega_0} \int_0^t dt' \big[\partial_X \omega_p^2(X,T) \big] _{X = x+c(t-t')}^{T = t'}.
\end{multline}

Equations~(\ref{frequp1}) and (\ref{kup1}) demonstrate that the change of laser frequency and wave vector is determined by the total temporal and spatial change of plasma density, respectively. In the limit of instantaneous plasma creation, the upshift of frequency and wave vector after interaction reduce to a simple form $\Delta\omega \equiv {\omega-\omega_0} = \omega_p^2/(2\omega_0)$ and $\Delta k \equiv {k - k_0} = \omega_p^2/(2c\omega_0)$. 
The relation of the instantaneous laser frequency and wave vector, $\Delta\omega = c\Delta k$, becomes very useful for interpreting our numerical simulation results: whereas experiments measure the laser frequency at a specific location, numerical simulations often more conveniently output the laser wave vector at a specific time. Equations.~(\ref{frequp1}) and (\ref{kup1}) provides a definite relation to transform the laser wave vector upshift into frequency upshift during the QED cascade.

\begin{figure}[th]
	\centering
	\includegraphics[width=0.8\linewidth,valign=t]{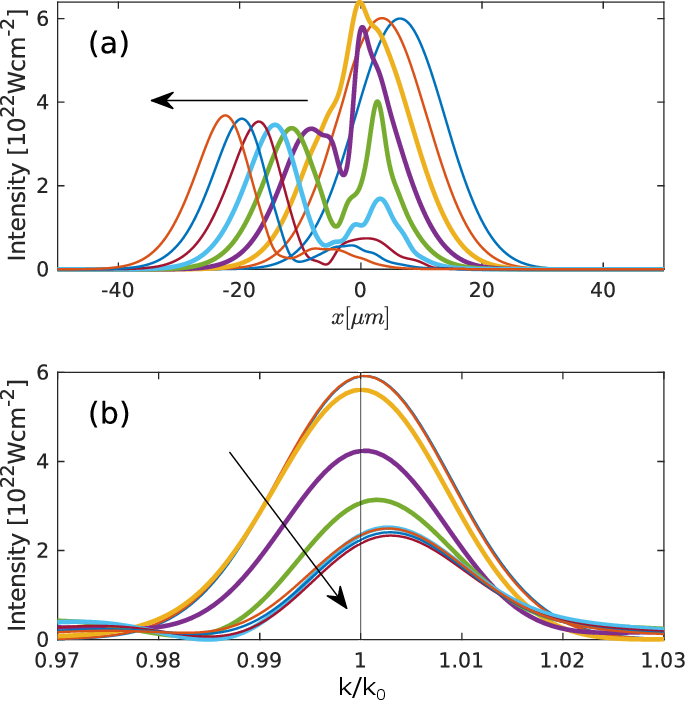}
	\caption{(a) The laser pulse intensity envelopes between $t=\unit[0.16]{ps}$ and $\unit[0.24]{ps}$ with an increase of time in the direction of the arrow. (b) The laser intensity spectra at the corresponding times. } 
	\label{enve}
\end{figure}

For the simulation under consideration, the peak value of $n_p/\gamma$ corresponds to $6.7\%$ of the critical plasma density at rest $n_c\approx 1.71\times10^{21}\,\mathrm{cm}^{-3}$ of the drive laser.  
Accordingly, a laser frequency upshift is observed in the intensity spectra displayed in Fig.~\ref{enve}. Figure~\ref{enve}(a) shows the laser intensity in its propagation axis $y=z=0$ with each curve corresponding to  from $t=\unit[0.16]{ps}$ to $\unit[0.24]{ps}$, respectively, in the direction of the arrow. Figure~\ref{enve}(b) shows the corresponding intensity spectra by Fourier transformation. The peaks of the spectra before and after the collision reveals a wave vector upshift  $\Delta k/k_0= 0.2\%$. Since the laser pulse propagates against the  pair plasma, its  wave vector spectrum becomes equivalent to the frequency spectrum after the collision: $0.2\%$ is also the laser frequency upshift. 
This finite frequency upshift is caused by the small fraction of laser overlap with the electron beam. Specifically, the frequency-upconverted photons are confined to a small region, whereas the majority of laser photons are not upconverted.

The oscillatory motion of the high density pairs absorbs a significant amount of laser energy. Combined with the QED process, it causes a decrease of laser peak intensity, which can be observed in Fig.~\ref{enve}. We highlight this process between $t=\unit[0.18]{ps}$ and $\unit[0.21]{ps}$ as thick curves in Fig.~\ref{enve}(a). This period corresponds to when the pair parameter $n_p/\gamma$ approaches its peak value, as can be seen from Fig.~\ref{evo}(b). Actually, the pairs, after absorbing the laser energy,  radiate to the whole space. It is revealed in Fig.~\ref{enve}(a) as splitting of the laser peak when the pairs are generated and laser frequency is upshifted. While the main laser peak continues to propagate to the $-x$ direction, a second peak is developed at $t=\unit[0.19]{ps}$ (thick purple curve) and propagates towards the $+x$ direction. 

Due to the small volume of the pairs, they emit a point-source-like radiation, as shown in Fig.~\ref{planes}(c). The large radiation angle can actually be used advantageously for experimental detection: It can be captured by an optical detector installed away from the path of the laser beam. The radiation is near the laser frequency and is hence easily distinguished from the high energy gamma photons.

\begin{figure}[th]
	\centering
	\includegraphics[width=\linewidth,valign=t]{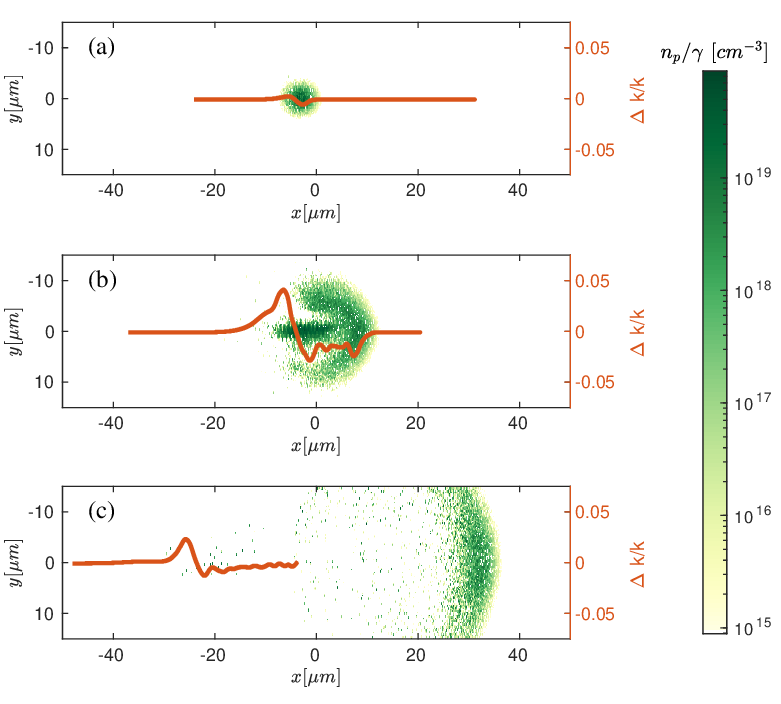}
	\caption{The pair parameter $n_p/\gamma$ and the relative laser wave number change $\Delta k/k$. The snapshots are taken at $\unit[0.16]{ps}$ (a), $\unit[0.2]{ps}$ (b), and $\unit[0.28]{ps}$ (c), respectively.} 
	\label{upshift}
\end{figure}

Much higher laser frequency upshift can be obtained when focusing on the region where pairs are quickly created. Such a laser spectrogram is typically obtained in experiments using techniques like frequency-resolved optical gating~\cite{FROG} or spectral shear interferometry for direct electric field reconstruction~\cite{SPIDER}. Numerically, we conduct a  wavelet transform of the laser pulse and obtain precisely the laser photon wave vectors at different pulse positions plotted as red curves in  Fig.~\ref{upshift}. We also plot the pair particle density in the $z=0$ plane to demonstrate the correlation of pair plasma creation and laser wave vector upshift. Figure~\ref{upshift}(a) shows that the wave vector spectrum becomes chirped immediately at the region of plasma creation near $x=0$. The wave vector chirps up in the front of the interaction region and chirps down in the tail, which agrees with Eq.~(\ref{kup1}). As the pair density increases and the interaction continues, the amplitude of chirp grows, as seen in Fig.~\ref{upshift}(b). The chirped region propagates along the laser direction [Fig.~\ref{upshift}(c)] and gets separated with the pair plasma. Thus, it can eventually be collected by a detector and reveals as a chirped frequency spectrum. The maximum photon frequency shift reaches $\Delta \omega/\omega_0= 2.4\%$.

\begin{figure}[ht]
	\centering
	\includegraphics[width=\linewidth,valign=t]{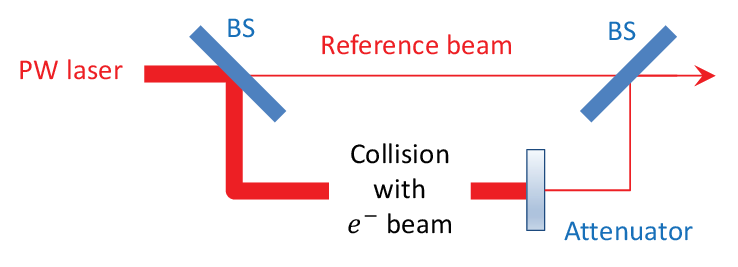}
	\caption{Interferometer setup for homodyne detection of the change in laser profile. A weak reference beam is split from the strong PW laser using a beam splitter (BS) with a large reflection ratio. The same reference beam is then combined with the attenuated post-interaction PW laser to produce the interference signal. } 
	\label{interf1}
\end{figure}

\begin{figure}[h]
	\centering
	\includegraphics[width=\linewidth,valign=t]{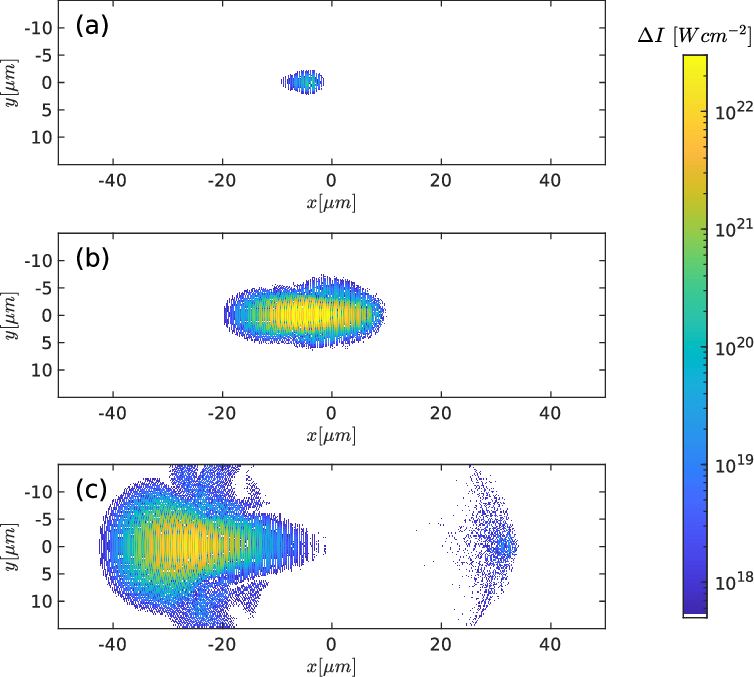}
	\caption{The homodyne signal which shows the change of laser beam intensity profile at the $z=0$ cross section. The snapshots are taken at $\unit[0.16]{ps}$ (a), $\unit[0.2]{ps}$ (b), and $\unit[0.28]{ps}$ (c), respectively.} 
	\label{interf2}
\end{figure}

The small disturbance in laser phase and intensity can be precisely measured with an interferometer. As shown in Fig.~\ref{interf1}, the strong laser is first sent through a beam splitter with a large reflection ratio. A small fraction of the laser pulse is split to serve as a reference beam, whose electric field can be denoted as $E_r = E_{r0}(\bm{r},t) e^{-i\omega_0 t} + c.c.$ The strong laser pulse, after interacting with the electron beam, becomes $E = E_{0}(\bm{r},t) e^{-i\omega_0 t + i\varphi} + c.c.$ . Here, $\varphi\equiv \int_0^t \Delta\omega dt'$  represents the  accumulated local phase change and $E_{0}(\bm{r},t)$ denotes the new envelope. The pulse is then attenuated to the same amplitude with the reference beam before they are combined  through another beam splitter. The interference signal, called a homodyne signal, is 
\begin{equation} \label{homodyne}
	\Delta I = \frac{1}{T} \int_0^T \frac{c\varepsilon_0}{2} |E_r - E|^2 dt, 
\end{equation}
where the negative sign arises from the double reflection of the PW laser and the signal is averaged through an optical cycle $T=2\pi/\omega_0$ to model the slow response time of the photo detector. The homodyne signal is sensitive to both the laser phase fluctuation and envelope change. For only a small phase fluctuation $\varphi \ll 2\pi$, the homodyne signal is proportional to the accumulated phase $\varphi$, \ie
\begin{equation}\label{homodyne2}
	\Delta I \propto [1-\cos(\varphi)]  I_{r0} \approx \varphi I_{r0},
\end{equation}
where $I_{r0}$ is the intensity of the reference beam. Since $\varphi=\int_0^t \Delta\omega dt'$, one can find out the frequency shift $\Delta\omega$ through $\Delta I/I_{r0}$. Note that the proportional relation in Eq.~(\ref{homodyne2}) only holds for $\Delta I \ll I_{r0}$. 

Numerically, we show the homodyne signal $ (c\varepsilon_0/2)|E_r - E|^2$ at three different snapshots in  Fig.~\ref{interf2} assuming instantaneous detector response time. Slower detectors will not detect the wavelength-scale fringes but the intensity envelope will be the same. The reference beam is obtained via a separate simulation of the same laser beam without encountering the electron beam. The interference signifies the change of electromagnetic field. 
The radiation pattern in Fig.~\ref{interf2}(c) consists of two types of signals, including a pattern that propagates along the laser direction and a pattern that radiates to the whole space. The propagation radiation pattern is the result of interference of phase-shifted post-interaction beam and the reference beam. The latter radiation pattern is the point-source-like pair emission caused by the finite pair plasma size. It confirms that the radiation starts at the location where pairs are generated and then expands to the whole space.

\section{Scaling of the laser frequency upshfit}

Since upconversion of laser frequency is determined by the pair plasma frequency, it provides an unambiguous signature of collective plasma effects in beam-driven QED cascades. In the previous section, our 3D PIC simulations demonstrate the collective pair plasma effects during pair creation and energy decay, and show how the plasma signature is imprinted in the colliding laser. For purposes of illustration, the collision uses a $\unit[24]{PW}$ laser pulse and a $\unit[1]{nC}$ electron beam at $\unit[300]{GeV}$. But can existing technology produce sufficiently high density pair plasma to exhibit observable collective effects?  In this section, we answer the question by finding how the amount of frequency upshift scales with different parameters of the laser and electron beam.

It is made clear in Eqs.~(\ref{eqA}) and (\ref{frequp0}) that the magnitude of pair plasma radiation is determined by the pair plasma parameter $n_p/\gamma$, \ie plasma density divided by pair energy. The spatial profile of the radiation depends on the pair size at the time of pair creation. High pair density is achieved through high input electron beam density $n_e$ and large pair multiplication factor $\widetilde{\chi}$, which, according to Eq.~(\ref{multip}), are proportional to the laser amplitude $a_0$, laser frequency $\omega_0$, and electron beam energy $\gamma_0 m_ec^2$. 

For exhibiting collective effects, an equally important parameter is the pair energy or Lorentz factor. In a QED cascade, the pair energy decreases as a result of radiation recoil and ponderomotive potential of the laser field, as we explained in Sec.~IV. The minimum pair energy is reached when the laser intensity meets the threshold for pair reflection, as shown in Eq.~(\ref{eqn:reflection}). Then, the pair motion becomes purely transverse and the pair Lorentz factor reduces to $a_0$, as shown in Eq.~(\ref{eqn:ruleofthumb}).
By combining \eqref{frequp0} with \eqref{eqn:ruleofthumb}, we find the frequency upshift
\begin{gather}\label{frequp}
\omega_f^2/ \omega_0^2  - 1 \sim \frac{\widetilde{\chi}_e n_e}{n_ca_0} \sim \gamma_0 \frac{\hbar\omega_0}{m_ec^2} \frac{n_e}{n_c}.
\end{gather}
This relation holds if the laser pulse is sufficiently long and intense such that the QED cascade fully develops and the pair plasma is eventually stopped and reflected.

The scaling relation [Eq.~(\ref{frequp})] is verified through a series of 1D QED-PIC simulations as reported in Ref.~\cite{Qu_QED2021}. The data show that  increasing either beam density or beam energy causes a linear increase of the created pair plasma density, whereas the final pair Lorentz factor remains constant at about $a_0$. They both results in a linear increase of the maximum frequency upshift. 

The effect of higher laser intensity has a threshold dependence.  When the laser intensity is below $3\times 10^{22} \,\mathrm{W cm}^{-2}$, the pair reflection condition [\eqref{eqn:reflection}] is not satisfied and the cascade does not saturate within the pulse duration, causing minimum laser frequency upshift. But above the threshold amplitude, the quantity $n_p/\gamma$ becomes independent of the laser intensity and the laser frequency upshift reaches its maximum value.  


\section{Collision of 3~PW laser and 30~G\lowercase{e}V electron beam}

\begin{figure}[b]
	\centering
	\includegraphics[width=\linewidth,valign=t]{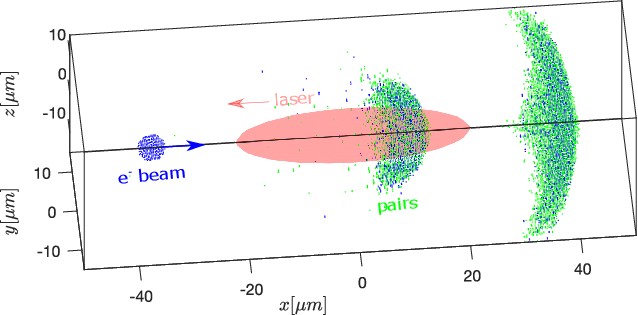}
	\caption{3D simulation of a $30\, \mathrm{GeV}$ (blue) electron beam colliding with a $3$~PW laser pulse (red). A  QED cascade creates a pair plasma (green) with an expanding volume. The electron beam itself also expands due to the laser ponderomotive force. } 
	\label{schem30}
\end{figure}

In the previous sections, we presented a clear numerical demonstration that the interplay between collective plasma and strong-field quantum effects leaves a characteristic imprint on the driving laser pulse. The upshift of the instantaneous laser frequency, according to the ``rule of thumb'' [\eqref{frequp}],  becomes experimentally observable by combining a multi-GeV class electron beam with density above ${{\sim}10^{20}}\, \mathrm{cm}^{-3}$ and a laser at ${\sim}\unit[10^{22}]{Wcm^{-2}}$ intensity. 
In principle, such beam-driven QED cascades could be initiated with electron beams
obtained from either a linear accelerator or laser wakefield acceleration (LWFA) at all-optical laser facilities. But linear accelerators benefit from its much higher total charge number at nC-level compared with pC-level obtained from reported LWFA accelerated electrons~\cite{PRL_LWFA2019}. The required electron beam density and beam energy can be obtained with only a moderate upgrade of existing facilities, e.g., SLAC's FACET-II~\cite{FACET-II}.

\begin{figure}[t]
	\centering
	\includegraphics[width=\linewidth,valign=t]{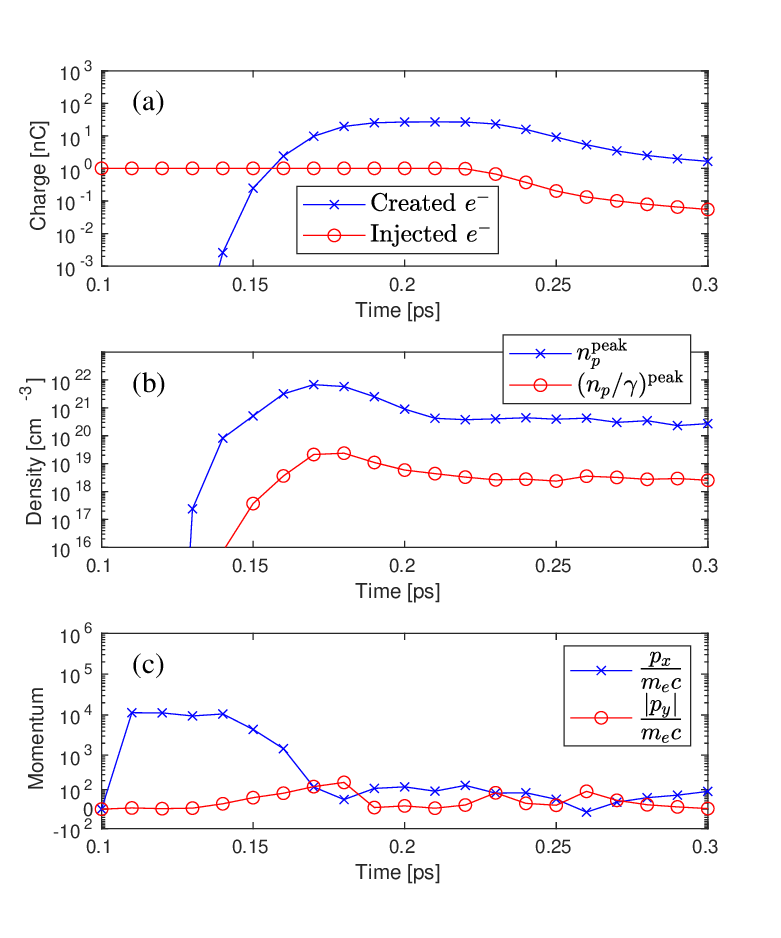}
	\caption{(a) Evolution of total charges of the injected $\unit[30]{GeV}$ electrons (red) and the created electrons (blue). (b) Evolution of the peak pair plasma density $n_p$ (blue), and the parameter $n_p/\gamma$ (red) which determines the laser frequency upshift. (c) Evolution of the pair particle momenta in the longitudinal (blue) and transverse (red) directions, normalized to $m_ec$. The local particle density $n_p$, Lorentz factor $\gamma$, and momentum $p_{x,y}$ denote their respective averaged values in a single simulation cell. } 
	\label{evo30}
\end{figure}

\begin{figure}[t]
	\centering
	\includegraphics[width=\linewidth,valign=t]{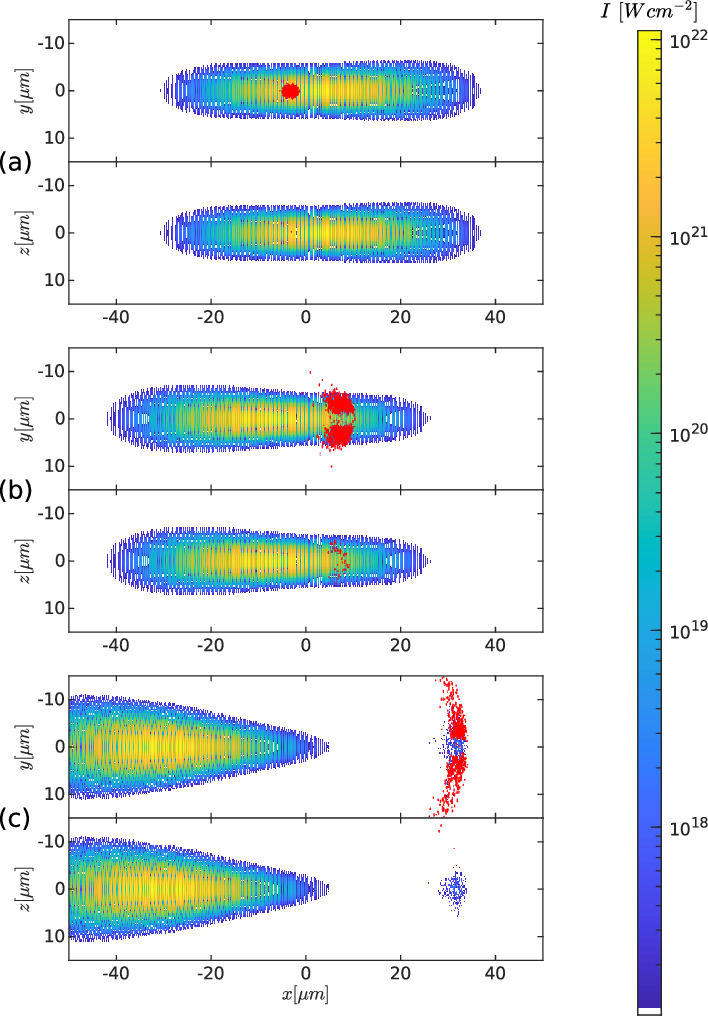}
	\caption{The two panels of each subplot show the laser profiles at the $z=0$ cross section and $y=0$ cross section, respectively. The red dots show the regions of $n_p/\gamma> \unit[1\times{}10^{19}]{{cm}^{-3}}$ at the corresponding planes. The snapshots are taken at $\unit[0.16]{ps}$ (a), $\unit[0.18]{ps}$ (b), and $\unit[0.24]{ps}$ (c), respectively.} 
	\label{planes30}
\end{figure}

We conduct such 3D QED-PIC simulations to show the prospect. Figure~\ref{schem30} illustrates the simulation of a $50$~fs-duration, $2.5~\mu$m-waist, $3$~PW ($\unit[3 \times 10^{22}]{Wcm^{-2}}$) laser pulse~\cite{danson_2019} colliding with a $\unit[1]{nC}$, $30\, \mathrm{GeV}$, ${4\times10^{20}}\, \mathrm{cm}^{-3}$ electron beam~\cite{FACET-II}. Other numerical parameters are identical to the last 3D simulations. The generated pair plasma and the radiation are shown in Figs.~\ref{evo30} and \ref{planes30}. Figure~\ref{evo30}(a) and (b) show that the created electron-positron pair plasma reaches a total charge number of $26\,\mathrm{nC}$ and peak density of ${6.8\times10^{21}}\, \mathrm{cm}^{-3}$. The red dots in Fig.~\ref{planes30} show that the tightly focused laser punches a hole in the pair plasma and pushes the pairs away from the propagation axis. The created pairs begin to escape the simulation box from $t=\unit[0.23]{ps}$ causing a decrease of total charge number as seen in Fig.~\ref{evo30}(a). 

With a lower energy, the injected electron beam has a much lower relativistic mass. The electrons are thus expelled by the strong laser ponderomotive force and they expand with the created pair plasma. This is seen in Fig.~\ref{schem30} where the blue dots mostly overlap with the green shades whereas the blue dots remain a sphere shape in Fig.~\ref{sch}. The expanding electron beam begins to leave the simulation box, causing a decreasing total charge from $t=\unit[0.23]{ps}$ as seen in Fig.~\ref{evo30}(a).

Since the laser intensity meets the threshold value $a_{0,\mathrm{th}}$ for particle reflection, some pairs reverse their propagation direction. This is confirmed in Fig.~\ref{evo30}(c) as the pair longitudinal momentum changes sign. The parameter $n_p/\gamma$ reaches ${2.4\times10^{19}} \, \mathrm{cm}^{-3}$, corresponding to $1.4\%$ of the critical density of a $0.8\mu\mathrm{m}$ laser. 
\begin{figure}[t]
	\centering
	\includegraphics[width=\linewidth,valign=t]{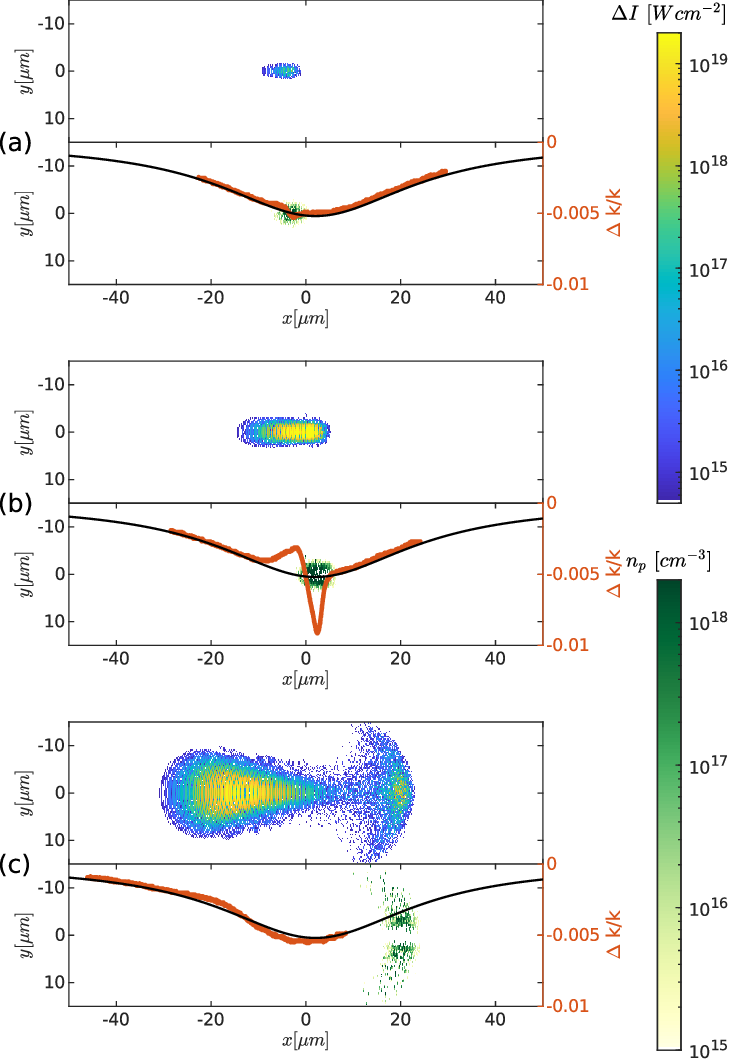}
	\caption{The top panel of each subplot shows the homodyne signal at the $z=0$ cross section. The bottom panel shows the pair parameter $n_p/\gamma$ at the $z=0$ cross section.  The red and black curves show the instantaneous wave vectors of the laser field at $y=z=0$ with and without encountering the electron beam, respectively.		The snapshots are taken at $\unit[0.16]{ps}$ (a), $\unit[0.18]{ps}$ (b), and $\unit[0.24]{ps}$ (c), respectively.} 
	\label{upshift30}
\end{figure}

We obtain the instantaneous wave vectors through a wavelet transform of the laser electric field at $y=z=0$ and plot them as red curves in Fig.~\ref{upshift30}. For comparison, we also plot the wave vectors of the same laser without encountering an electron beam as black curves. Due to the tight focus, the laser has a short Rayleigh length, so the Gouy phase induces a down chirp of the wave vectors near the focal point. The red curves clearly show a laser wave vector upshift in the region of pair plasma creation. The maximum wave vector upshift reaches a value of $\sim0.2\%$. The value, however, decreases as the pair plasma expands.

In the top panels of Fig.~\ref{upshift30}, we  plot the homodyne signal $(c\varepsilon_0/2) |E-E_r| $. Strong signals exhibit in the regions of pair plasma creation. The plots show that the interference signal immediately appears when the pairs are initially created. As the pair density grows, the signal reaches a maximum intensity of $\unit[1.2\times10^{20}]{Wcm^{-2}}$,which can be easily detected in an experiment. At $t\geq \unit[0.18]{ps}$, the homodyne signal intensity increases by $~\unit[5\times10^{19}]{Wcm^{-2}}$ within a single laser period indicating a $~0.16\%$ laser wave vector upshift.

\section{Summary and Discussion}

In couching the demonstration of collective plasma effects in the QED plasma regime as a  joint production-observation problem, we showed, using PIC simulations, how collective effects may in fact be both produced and observed in high-density electron-positron pair plasmas through strong-field QED cascades using existing laser and electron beam technologies. Both high pair density $n_p$ and low pair Lorentz factor $\gamma$ are shown to be equally important for exhibiting the strong collective plasma effects. A large pair plasma frequency can be achieved with a dense and high-energy electron beam. Higher laser intensities, however, does not always benefit  laser frequency upshift.
A higher laser intensity can effectively increase the pair plasma frequency only below the threshold amplitude $a_{0,\mathrm{th}}\gtrsim 100$ ($I_{0,\mathrm{th}}\gtrsim \unit[10^{22}]{Wcm^{-2}}$). With pairs being stopped, higher laser intensities above the threshold drive larger transverse momentum, and thus suppress the growth of plasma frequency. 

We specifically considered two sets of 3D PIC simulations, including one set with ideal parameters and the other set with existing technologies. 
The ideal parameter set uses a $\unit[24]{PW}$ $\lambda=\unit[0.8]{\mu m}$ laser with a waist of $\unit[5]{\mu m}$, duration of $\unit[50]{fs}$, and peak intensity of $\unit[6\times10^{22}]{Wcm^{-2}}$ (corresponding to $a_0\approx170$). Combined with a $\unit[300]{GeV}$, $\unit[1]{nC}$, $\unit[4\times10^{20}]{cm^{-3}}$ peak density  electron beam, it creates a $\unit[139]{nC}$ plasma with peak density of $\unit[3.28\times10^{22}]{cm^{-3}}$, which is $82$ times higher than the injected electron beam. The pair parameter $n_p/\gamma$ reaches a peak value of $\unit[2.67\times10^{20}]{cm^{-3}}$, which is $6.7\%$ of the critical density of the laser. The simulation shows a laser frequency upshift of $2.4\%$ at the output and a maximum wave vector upshift of  $4.8\%$ during the collision.

We further demonstrated that the collective QED plasma signature can be observed with state-of-the-art technologies. Even a $\unit[30]{GeV}$ electron beam and a $\unit[3]{PW}$ $\unit[2.5]{\mu m}$-waist laser (other parameters are identity to the previous 3D PIC simulation) can create a pair plasma of $n_p/\gamma = \unit[2.67\times10^{20}]{cm^{-3}}$ and induce a $0.2\%$ laser frequency upshift. It suggests how hitherto unobserved collective effects can be probed with existing state-of-the-art laser and beam technologies, offering a strong argument for colocating these technologies.

What emerges from these simulations is a picture of how the co-location~\cite{meuren2020seminal, meuren2021mp3} of a dense, $\unit[30]{GeV}$ electron beam with a $\unit[3-10]{PW}$ optical laser enables us to reach the QED plasma regime at substantially lower laser intensities ($\unit[10^{22}] {Wcm^{-2}}$) than previously thought possible. Whereas it has been well known that QED cascades can be studied in electron-beam laser collisions\cite{sokolov_pair_2010}, our simulations highlight the importance of the electron beam density, showing that the beam-compression techniques developed in the context of FACET-II are indeed sufficient to describe the interplay between collective plasma and strong-field quantum effects with a laser-electron-beam setup. 
In fact, compared to all-optical methods of reaching the QED regime, the use of lower laser intensities reduces the particle mass shift, thereby remarkably making the QED collective effects easier to observe. 

It can be imagined that further, possibly more distinct, signatures of the collective effects recognized here might be obtained through variations of the laser or the beam parameters envisioned here.  However, what is clear is that the beam-laser collision setup, together with the methodology of witnessing collective effects through laser frequency shifts, as envisioned here, already opens up possibilities for the near-term studying of the QED plasma regime with presently available technology.

\begin{acknowledgments}
	This work was supported by DOE Grants DE-NA0003871 and DE-AC02-76SF00515. 
\end{acknowledgments}  

\bibliography{Upshift}

\end{document}